\documentclass[aip,amsmath,a4paper, floatfix]{revtex4-1}
\usepackage{graphicx}
\usepackage{amsmath}
\usepackage{amsfonts}
\usepackage{amssymb}
\usepackage{color}
\usepackage{dcolumn}

\usepackage{bm}

\newcommand{\degree}{\ensuremath{^\circ}}

\usepackage{hyperref}

\begin{document}

\title{In-plane electronic confinement in superconducting LaAlO$_3$/SrTiO$_3$ nanostructures}

\author{D. Stornaiuolo}
\email{daniela.stornaiuolo@unige.ch}
\author{S. Gariglio}
\affiliation{DPMC - University of Geneva, 24 quai Ernest-Ansermet, CH-1211 Geneva, Switzerland}
\author{N.J.G. Couto}
\affiliation{DPMC - University of Geneva, 24 quai Ernest-Ansermet, CH-1211 Geneva, Switzerland}
\affiliation{GAP - University of Geneva, 24 quai Ernest-Ansermet, CH-1211 Geneva, Switzerland}
\author{A. F\^ete}
\author{A. D. Caviglia}
\altaffiliation{now at Max-Planck Research Group for Structural Dynamics-Center for Free Electron Laser Science, University of Hamburg, Notkestrasse 85, 22607 Hamburg, Germany}
\author{G. Seyfarth}
\altaffiliation{now at LNCMI-G, CNRS, 25 rue des Martyrs, 38042 GRENOBLE cedex 9, France }
\author{D. Jaccard}
\affiliation{DPMC - University of Geneva, 24 quai Ernest-Ansermet, CH-1211 Geneva, Switzerland}
\author{A.F. Morpurgo}
\affiliation{DPMC - University of Geneva, 24 quai Ernest-Ansermet, CH-1211 Geneva, Switzerland}
\affiliation{GAP - University of Geneva, 24 quai Ernest-Ansermet, CH-1211 Geneva, Switzerland}
\author{ J.-M. Triscone}
\affiliation{DPMC - University of Geneva, 24 quai Ernest-Ansermet, CH-1211 Geneva, Switzerland}

\date{\today}

\begin{abstract}
 
We describe the transport properties of mesoscopic devices based on the two
dimensional electron gas (2DEG) present at the LaAlO$_3$/SrTiO$_3$ interface. 
Bridges with lateral dimensions down to 500~nm were realized using
electron beam lithography. Their detailed characterization shows that processing and confinement do not alter the transport parameters of the 2DEG. 
The devices exhibit superconducting behavior tunable by electric field effect. In the normal state, we measured universal conductance
fluctuations, signature of phase-coherent transport in small structures.
The achievement of reliable lateral confinement of
the 2DEG opens the way to the realization of quantum electronic devices at
the LaAlO$_3$/SrTiO$_3$ interface.

\end{abstract}
\maketitle

Since its discovery,\cite{ohtomo_wang_Nature04} the two
dimensional electron gas present at the interface between the two
insulators LaAlO$_3$ and SrTiO$_3$ (LAO/STO) has been the subject of intense
study.
This large effort has brought to light a variety of
physical phenomena exhibited at this interface, making it one
of the most fascinating systems in the field of oxide electronics.\cite{mannhart,zubko}
The 2DEG
undergoes a superconducting transition with a maximum critical temperature of
$\approx$300~mK, \cite{reyren_science07} which can be gate tuned,
allowing the superconducting
state to be switched on and off.\cite{caviglia_nature08,bell_PRL09}
Magnetotransport measurements have shown that, concomitantly with the
superconducting state, a large Rashba spin-orbit interaction sets
in.\cite{caviglia_prlso,benshalom_so}
The out-of-plane confinement of the electron gas, estimated to be
on the scale of a few nanometres,\cite{basletic_nature08,reyren_APL09} gives rise
to two-dimensional electronic states, as recently observed via angle-dependent
Shubnikov-de Haas
oscillations. \cite{caviglia_SdH,benshalom}
\\
The physical parameters obtained from the electronic band structure
and the transport analyses are radically different from those of the 2DEGs found in classical
semiconductor
heterostructures, making this oxide interface particularly interesting for the
study of mesoscopic effects.\cite{gariglio_nano, mannhart}
For instance, the Fermi and the Rashba
spin-splitting energies are comparable ($\sim$10~meV). Additionally, in the
presence of a magnetic field, the large effective mass makes the Zeeman energy
larger than the Landau level splitting, a situation markedly different from the
one found in III - V heterostructures.
The breaking of inversion symmetry could also be at the origin of a
mixing of a singlet and triplet superconducting pairing state, which can be
investigated using tunnel junctions in spectroscopy experiments.\cite{kaur,liu} 
The realization of nanostructures is fundamental in order to explore these unconventional regimes. Two nanofabrication techniques have been applied up to now to this system.
In the first one, the voltage-biased metallic tip of an atomic force microscope is
scanned over the
surface of a 3 unit cell (u.c.) thick LAO film, locally inducing a conducting
channel at the interface. Metallic lines as narrow as a few nanometres
have been created and
measured.\cite{levy_1,levy_2,xie} The second approach relies on the use of
electron beam lithography,\cite{mannhart_APL06} which allows greater flexibility for the
realization of devices with complex geometries and longer lifetime. However, the evolution of the transport parameters with the bridge size and the possible effects of the 
fabrication process on the 2DEG's normal and superconducting properties need to be fully addressed in order to exploit the full potential of LAO/STO for nano-electronics.
\\
In this letter, we describe a systematic study of the transport properties of
LAO/STO-based nanostructures realized by electron beam
lithography. We show that the mobility and carrier density 
of bridges with lateral dimensions down to 500~nm are unaffected by the
fabrication process and are comparable with those of large area devices. At
low temperature, these bridges exhibit superconducting behavior with a critical current
that can be modulated using the field effect.
For magnetic fields large enough to suppress superconductivity, magnetoresistance measurements display
universal conductance fluctuations (UCF). 
As UCF stem from phase-coherent
transport,\cite{caviglia_nature08,caviglia_prlso,dagan_PRB10}
their observation indicates that the devices fabricated with the
technique described in this paper allow the mesoscopic transport
regime to be accessed.

\begin{figure}[t]
\centering
\includegraphics[width=9.0 cm, height=7.0 cm]{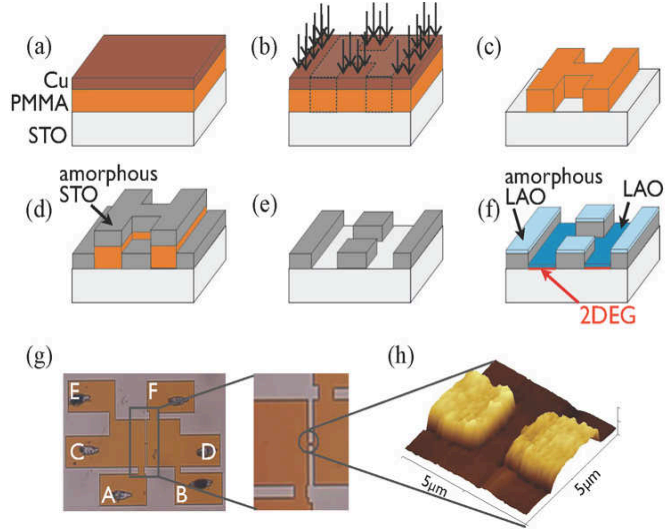} 
\caption{(Color online) Sketch of the fabrication process:
(a) deposition of PMMA and Cu layers, (b) e-beam exposure, (c) development of PMMA (after the Cu wet etch), (d) deposition of amorphous STO and (e) lift off, (f) deposition of the LAO film. In
panel (g) an optical microscope image of one of the devices is shown before the
amorphous STO lift-off procedure
(full scale: 1.5~mm$\times$1.2~mm). Panel
(h) shows an atomic force microscope image of the bridge area.
Polycrystalline/amorphous LAO, grown on the amorphous STO, has a lighter color,
while the epitaxial LAO has a darker one. The 2DEG is created only below the epitaxial LAO.}
\end{figure}

The steps for the fabrication of nanoscale bridges are illustrated in Figure 1.
The process avoids an ion milling step, which could damage
the STO crystal with, for instance, the creation of oxygen vacancies.
\cite{henrich}
A template with the desired pattern is realized on the substrates by depositing an
amorphous STO layer before the epitaxial growth of the LAO thin film.
To this purpose, we pattern a resist mask on the bare
STO substrate (Figures 1(a), (b) and (c)) using an electron beam lithography
procedure optimized for insulating materials. 
Specifically, prior to exposure, we cover the PMMA resist with a thin film of copper (7~nm, see Figure 1(a))\cite{PMMA}. This metallic film acts as a charge
dispersion layer, reducing accumulation of surface charges on the STO 
insulating substrate. It has a minimal influence on the pattern 
resolution, while increases slightly the exposure dose required for PMMA.\cite{Cu_overlayer} After exposure, the 
copper layer is removed by wet etching using a solution of FeCl in 
water, then the PMMA is developed using a diluted methyl isobutyl 
ketone (MIBK) solution, revealing the pattern.
After development, we deposit a thin layer (15~nm) of amorphous STO (Figure
1(d)), before removing the resist
with a lift-off procedure. Thus we obtain a template substrate with the desired
pattern made of an amorphous STO layer (Figure 1(e)).\cite{banerjee} Finally,
an epitaxial 10 u.c.-thick LAO film is deposited using pulsed laser
deposition (Figure 1(f)).\cite{cancellieri} The film is grown at 800\degree C in
10$^{-4}$~mbar of
oxygen. The KrF excimer laser fluency is 0.6~J/cm$^2$ with a repetition rate of
1~Hz. Immediately after deposition, the sample is annealed in oxygen: we fill
the deposition chamber with 200 mbar of oxygen and keep the
sample temperature at 520\degree C for 1~hour.
The sample is then slowly cooled down to room temperature in the same
oxygen atmosphere.\cite{aSTO} The growth process is monitored {\it in
situ} using reflection
high energy electron diffraction (RHEED), which shows, for all the samples, a
layer-by-layer growth mode.
An atomic force microscope image of an 800~nm-wide and 2~$\mu$m-long
bridge realized with this technique is shown in Fig. 1(h). 

We have fabricated several tens of bridges of different width, with good
reproducibility of the transport
parameters.
Using the layout shown in Fig. 1(g), we are able to measure the
properties of both the bridges and of the larger area adjoining each of them,
using a four-point DC technique.
We inject current through contacts C-D and measure the voltage drop
between A and F to probe the bridges, and between B and F (or A
and E) to probe the larger areas.

\begin{figure}[t]
\centering
\includegraphics[width=5.5 cm, height=8.15 cm]{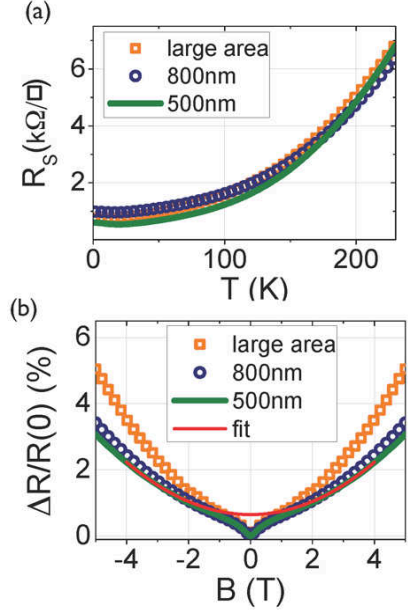} 
\caption{(Color online) Transport properties of the mesoscopic devices. Panel
(a) shows typical $R_S(T)$ curves of 500~nm (green line) and 800~nm (blue
circles) bridges and of a large area (300$\mu m \times 300~\mu m$)
(orange
squares). Panel (b) displays the relative magnetoresistance, measured at
1.5~K, for the same three devices. The magnetic field was applied perpendicular to the sample. The solid red line is an example of the
quadratic fit used to extract the carrier mobilities.}
\end{figure}

Figure 2(a) shows the metallic behavior of two devices, 500~nm (green line) and
800~nm (blue circles) wide, both 2~$\mu$m-long, and of a large area of
dimensions 300~$\mu$m$\times$300~$\mu$m (orange squares).
Figure 3(a) summarizes the sheet resistance $R_S$ values measured at 1.5~K for
devices of
different size realized on the same chip and for the large areas adjoining each of them. Figure 2(b)
displays the magnetoresistance $\Delta R / R(0)= (R(B)-R(0))/R(0)$ of the same
devices, measured at 1.5~K. Fitting the quadratic behavior of the
magnetoresistance with the relation \cite{pearson,kuchar}
$\Delta R / R(0)=(\mu B)^2$ (see for example the fit referring to the
500~nm bridge in Figure 2(b)), we estimate the mobility $\mu$ of the
charge carriers for channels of different width as well as for the large areas. These data are plotted in Figure 3(b). From these
measurements, we can also estimate the number of
carriers in the bridge $n_{2D}=1/(\mu R_S e)$ ($e$ is the electron charge).
As shown in Figure 3(c), $n_{2D}\sim 3\cdot 10^{13}cm^{-2}$ (green diamonds), a
typical value for LAO/STO interfaces.
The analysis of the Hall data measured directly across the bridge and in the
large areas yields comparable carrier densities, as can be observed in the
same Figure.
We observe that the values shown in Figure 3 change by a factor of 3 for
channel widths spanning almost two orders of magnitude from 500~nm to
10~$\mu$m,\cite{afm} as well as for the large areas of different devices.
Fluctuations of this amplitude in the sheet resistance and mobility are fully
compatible with variations observed from sample to sample, as reported
in the literature by different groups for the same growth
conditions.
These results allow us to affirm that processing
and confinement down to the sub-micron scale do not alter the
carrier population and the transport properties, which appear to be homogeneous in the different areas of the
sample. 

\begin{figure}[t]
\centering
\includegraphics[width=6.5 cm, height=9.9 cm]{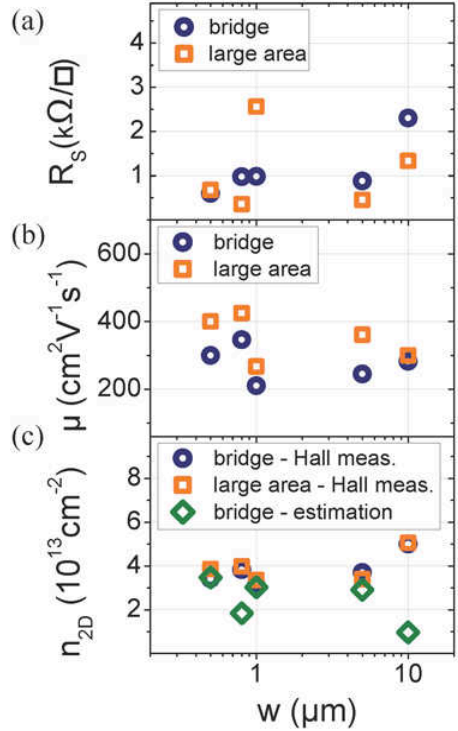} 
\caption{(Color online) Transport parameters measured at 1.5~K of bridges of different width (blue circles) and of the adjoining large areas (orange squares): sheet resistance (a), mobility (b) and number of carriers (c). The green diamonds in panel (c) are the number of carriers estimated using the formula $n_{2D}=1/(\mu R_S e)$ with $\mu$ and $R_S$ from panels (a) and (b).}
\end{figure}

When cooled to millikevin temperatures, these devices exhibit superconducting behavior.
We measured voltage vs. current ($V-I$) characteristics at T=40~mK in a dilution
refrigerator equipped with
copper powder filters, using a four terminal configuration.
Figure 4(a) and (b) show $V-I$ plots of the 500~nm and 800~nm wide
devices revealing a superconducting state
for these LAO/STO structures in the sub-micron range.
The critical current densities are 150 and 430~$\mu$A/cm for the 500~nm, 800~nm bridges respectively. 
These values are in line with a previous work\cite{reyren_science07} where large-scale devices were measured.   Their scatter is of the same order of that observed in transport parameters in the normal
state (see Figure 3).
The critical current can be tuned applying a gate voltage. 
Figure 4 shows the modulation of the $V-I$ characteristics for the 500~nm and
800~nm devices obtained in a back-gate configuration.\cite{caviglia_nature08}
We observe that a complete suppression of the critical current occurs for
V$_g$=-100~V for the 500~nm device (panel (a)).
For the 800~nm device, the zero resistance state is lost only below V$_g$=-160~V
(panel b). 

In the normal state, the confinement of the 2DEG allows us to access the mesoscopic transport regime which, for phase coherent transport, manifests itself with the appearance of universal conductance fluctuations.\cite{lee_stone_PRB87, beenakker_review}
UCF stem from quantum interference of phase-coherent electron waves scattered by
impurities in a sample with dimensions $L$ comparable to the phase
coherence length $l_\varphi$. For $L=l_\varphi$ the amplitude of such
fluctuations is expected to have the universal value $e^2/h$.
Estimations of $l_\varphi$ in LAO/STO interfaces yield values in the range
of 100-200~nm at 1.5~K. \cite{caviglia_prlso,dagan_PRB10,dikin_PRL11}
Therefore our small bridges are expected to display UCF at low
temperatures, albeit with a reduced amplitude since their dimensions are still
larger than $l_\varphi$. 

\begin{figure}[t]
\centering
\includegraphics[width=5.5 cm, height=11 cm]{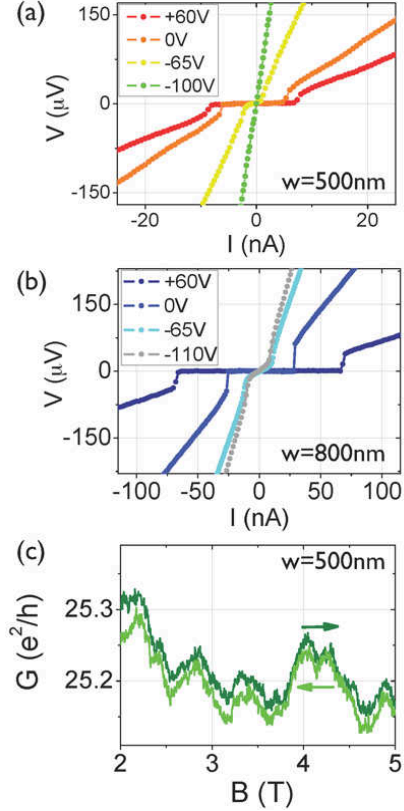}
\caption{(Color online) Low temperature characterization of the mesoscopic
bridges. Panels (a) and (b) show the gate modulation of the $V-I$ characteristics for the
500~nm and 800~nm wide devices respectively. Panel (c) shows magnetoconductance fluctuations measured for the 500~nm bridge. The conduction state was set using field effect. The two traces refer to opposite directions of the field sweep.
All the measurements presented in this figure
were carried out at 40~mK. }
\end{figure} 

Figure 4(c) shows a magnetoconductance trace of the 500~nm wide device
measured at 40~mK. The aperiodic fluctuations which can be seen
have all the
fingerprints of UCF. The two traces shown in the plot refer to the two sweep
directions
of the magnetic field, demonstrating the reproducibility of the
fluctuation pattern. The phase-coherence length $l_\varphi$ can be calculated from the typical 
average magnetic field spacing of the fluctuations $\Delta B_c$ via the
equation:\cite{lee_stone_PRB87,beenakker_review}
 $\Delta B_c = C \Phi_0/(l_\varphi^2)$, where $\Phi_0$ is the quantum of flux
and
$C$ is a constant of the order of unity.\cite{DBc}
This estimation leads to $l_\varphi \approx$ 110$\pm20$~nm.
If we now consider the amplitude of the fluctuations, we observe
that it is reduced with respect to the universal value. For the curve shown in
Figure 4(c) it amounts to about 8$\%$ of $e^2/h$. 
A reduction in the amplitude is indeed expected when the
mesoscopic channel has dimensions larger than $l_\varphi$. In this case, the
channel may be subdivided into coherent areas of
dimensions $l_\varphi\times l_\varphi$; the conductance fluctuations of
each of these areas will be of order $e^2/h$ and all the areas in the channel
will add incoherently, leading to classical self-averaging.
The resulting conductance fluctuations will be reduced by a factor proportional
to $\sqrt {wl/l_\varphi^2}$, with $w$ and $l$ the width and length of the
bridge respectively.\cite{lee_stone_PRB87}
Considering our geometry, our results are in good numerical agreement with this
argument.

In summary, we reported a detailed study of nanodevices based on the 2DEG at the LAO/STO interface. 
Mesoscopic bridges with widths down
to 500~nm were realized using electron beam lithography.
Possible damage to the STO substrate, such as oxygen losses,
were avoided by resorting to an amorphous STO template.
The detailed transport characterization reveals that processing
and confinement down to the sub-micron scale do not alter substantially the
carrier profile and that the samples are homogeneous down to the scale of our
smallest bridges. The confinement of the 2DEG in the in-plane direction results in universal
conductance fluctuations that were observed in magnetotransport at low
temperatures. The devices show superconducting behavior, with a critical
current tunable using field effect. The observation of a zero-resistance state in LAO/STO
nanodevices opens exciting perspectives for the study of quasi-one dimensional
superconductivity and for the realization of devices such as tunable Josephson
junctions. 
\\
We thank P.Zubko for careful reading of the manuscript and M. Lopes and S. C. M\"uller for their technical
assistance. This work was supported
by the Swiss National Science Foundation through the National Center of
Competence in Research, Materials
with Novel Electronic Properties, MaNEP, division II, and the European Union
through the project OxIDes.
\\


\begin{thebibliography}{99}
\bibitem{PMMA} We used two layers of PMMA, with 495~K and 950~K molecular
weight, both with a 3$\%$ solid content in anisole, for a total thickness of
$\sim$300nm.
The double layer was used in order to facilitate the subsequent lift-off step.
\bibitem{aSTO} At the end of our fabrication process, the interface between the
STO substrate and the amorphous STO layer is observed to be insulating.
\bibitem{afm} The lateral dimension of each bridge is confirmed using AFM
topography.
\bibitem{DBc} An exact estimation of the correlation field $\Delta B_c$ is given
by the full width at half maximum of the magneto conductance autocorrelation
function, which fully captures the characteristic length scales of a mesoscopic
system.
\end{thebibliography}
\end{document}